\documentstyle[preprint,revtex]{aps}
\begin{document}
\draft
\preprint{MSUTH 92/03}
\preprint{October, 1992}
\begin{title}
The effects of a strongly interacting Higgs sector \\
on $\gamma\,\gamma\rightarrow W_L^+\,W_L^-, Z_L^{\,0}\,Z_L^{\,0}$ scattering
\end{title}
\author{A. Abbasabadi}
\begin{instit}
Physical Sciences Department\\
Ferris State University, Big Rapids, Michigan 49307
\end{instit}
\author{David Bowser-Chao and Duane A. Dicus}
\begin{instit}
Center for Particle Physics\\
University of Texas, Austin,Texas 78712
\end{instit}
\author{Wayne W. Repko}
\begin{instit}
Department of Physics and Astronomy \\
Michigan State University, East Lansing, Michigan 48824
\end{instit}
\begin{abstract}
We study the effects of a strongly interacting Higgs sector on the amplitudes
for $\gamma\,\gamma\rightarrow W_L^+\,W_L^-, Z_L^{\,0}\,Z_L^{\,0}$ by
unitarizing the $W_L^+\,W_L^-,
Z_L^{\,0}\,Z_L^{\,0}, \gamma\,\gamma$ system using the $K$--matrix
technique. Unitarization produces substantial corrections to the cross sections
for gauge boson pair production by photon fusion when $m_H \agt 5-10\;$TeV.
\end{abstract}

\narrowtext
\section{INTRODUCTION}

The possibility of producing high-luminosity photon beams by backscattering
laser beams from high energy polarized electrons has prompted theoretical
investigations suggesting that the photon fusion processes
\begin{equation}
\gamma\,\gamma\rightarrow\,W_L^+\,W_L^-\,,\,Z_L^{\,0}\,Z_L^{\,0}\;,
\label{ggwwzz}
\end{equation}
can be used to probe the Higgs sector of the standard model \cite{boji} as
well as to explore the symmetry breaking sector of its non--linear Chiral
Lagrangian generalizations \cite{hr-m}. It has also been
shown that the photon fusion processes of Eq.(\ref{ggwwzz}) can be used
in the search for ultraheavy fermions at high energy hadron colliders
\cite{chan}. In either case, the $W_L^+\,W_L^-\,$ and $\,Z_L^{\,0}\,Z_L^{\,0}$
signals receive large contributions from one-loop diagrams containing
$W_L^{\pm}\,,\,Z_L^{\,0}\,$ and Higgs bosons, $H$ \cite{boji}.

In this paper, we examine the effect of a strongly interacting Higgs sector on
the photon fusion by unitarizing the s-wave amplitudes for $W_L^+\,W_L^-\,,
\,Z_L^{\,0}\,Z_L^{\,0}\;$ and $\gamma\,\gamma\,$ scattering using the
$K$-matrix
formalism \cite{repsuch,dicrepkm}. It is then possible to increase the
longitudinal coupling strength $\lambda = g^2\,m_H^2/8\,m_W^2$ by varying
$m_H$.
Additionally, the Higgs width is introduced in a manner which preserves the
Goldstone-boson equivalence theorem \cite{clt,lqt,cg}.

\section{UNITARIZATION}
\label{sec:unit}

We define the $3\,\times\,3\;,\;J\,=\,0\,$ $t$-matrix as
\begin{equation}
t_0 = \left(\;
\begin{array}{ccc}
a\,+\,b\;\;      &
\mathstrut\displaystyle{a\over\sqrt{2}}\;\;   &
\varepsilon\,\mathstrut\displaystyle{d\over\sqrt{2}} \\
\mathstrut\displaystyle{a\over\sqrt{2}}\;\; &
\mathstrut\displaystyle{1\over 2}\,a\,+\,b\;\; &
\varepsilon\,\mathstrut\displaystyle{a\over 2}      \\
\varepsilon\,\mathstrut\displaystyle{d\over\sqrt{2}}\;\; &
\varepsilon\,\mathstrut\displaystyle{a\over 2}\;\; &
\varepsilon^2\,\mathstrut\displaystyle{f\over 2}
\end{array}
\;\;\right)\;,
\end{equation}
where $\varepsilon = \alpha/2\,\pi\,$. The rows and columns of $t_0$ are
labeled
by $W_L^+\,W_L^-\,,\,Z_L^{\,0}\,Z_L^{\,0}/\sqrt{2}\;$ and $\gamma\,\gamma/
\sqrt{2}\;$, and the Born level partial wave projections $a$ and $b$ are given
by \cite{lqt}
\begin{eqnarray}
a &=\;&-{\lambda\over 8\,\pi}\,{s\over s - m_H^2}                \\
b &=\;&-{\lambda\over 8\,\pi}\,\left(1 - {m_H^2\over s}\,\ln(1 + {s\over
m_H^2})
\right)\;.
\end{eqnarray}
In Ref.\ \cite{boji}, it is shown that the one-loop $\gamma\,\gamma\rightarrow
\,Z_L^{\,0}\,Z_L^{\,0}$ amplitude is proportional to $a$ for
$m_W^2\rightarrow\,
0$. For the $\gamma\,\gamma\rightarrow\,W_L^+\,W_L^-\,$ case, the one-loop
correction to the s-wave amplitude is also real for $m_W^2\rightarrow\,0$,
since
the Born amplitude vanishes in this limit. However, explicit calculations show
that the Born term, though proportional to $m_W^2$, is not negligible for
interesting values of the $\gamma\,\gamma$ center of mass energy $\sqrt{s}$.
Consequently, the s-wave amplitude $d$ is taken to be
\begin{eqnarray}
d &=\;& 2\,\pi\,{m_W^2\over \beta_W\,s}\ln\left({1 + \beta_W\over 1 - \beta_W}
\right) - {\lambda\over 8\,\pi}\,{s\over s - m_H^2} \nonumber \\
  &   & \\
  &   &-{\lambda\over 8\,\pi}\left(1 + {2\,m_H^2\over s}\left({\rm Li_2}
(-{s\over m_H^2})  + \left(1 + {m_H^2\over s}\right)\ln(1 + {s\over m_H^2}) -1
\right)\right)\,.\nonumber
\end{eqnarray}
Since we are only interested in effects of order $\alpha$ and
$\alpha\,\lambda\,
$, we can ignore the photon-photon partial wave amplitude $f\,$ \cite{f}.
We also consider
the case of photons with helicity $\lambda_{\gamma} = 1\,$, which is
appropriate
for the case of backscattered laser photons.

Including terms through order $\alpha^2\,$, the eigenvectors of the matrix
$t_0$ are
\begin{eqnarray}
\xi_1 &=\;& \left({\sqrt{2}\over\sqrt{3}}(1 - \frac{1}{6}\,{y^2\over c^2} -
\frac{2}{9}\,{x\,y\over a\,c})\;,{1\over\sqrt{3}}(1 - \frac{1}{6}\,
{y^2\over c^2} + \frac{4}{9}\,{x\,y\over a\,c})\;,
{1\over\sqrt{3}}\,{y\over c}\right)\;, \\
\xi_2 &=\;& \left({1\over\sqrt{3}}(1 - \frac{1}{3}\,{x^2\over b^2} +
\frac{4}{9}\,{x\,y\over a\,b})\;,-{\sqrt{2}\over\sqrt{3}}(1 - \frac{1}{3}\,
{x^2\over b^2} - \frac{2}{9}\,{x\,y\over a\,b})\;,-{\sqrt{2}\over\sqrt{3}}\,
{x\over b}\right)\;,   \\
\xi_3 &=\;& \left({\sqrt{2}\over 3}\,({x\over b} - {y\over c})\,,
-(\frac{2}{3}{x\over b} + \frac{1}{3}{y\over c})\;,(1 - \frac{1}{6}\,{y^2\over
c^2} - \frac{1}{3}\,{x^2\over b^2})\right)\;,
\end{eqnarray}
where $c = \frac{3}{2}\,a + b\,$, $x = \frac{1}{2}\,\varepsilon\,(a - d)\,$ and
$y = \frac{1}{2}\,\varepsilon\,(a + 2\,d)\,$. It is easy to
check that if $U$ is the matrix
whose columns consist of $\xi_1\,$, $\xi_2\,$ and $\xi_3\,$, the product
$U^{\rm
T}\,t_0\,U\,$ is diagonal. The corresponding eigenamplitudes are
\begin{eqnarray}
\lambda_1 &=\;& c + \frac{1}{3}\,{y^2\over c}\;,    \\
\lambda_2 &=\;& b + \frac{2}{3}\,{x^2\over b}\;,    \\
\lambda_3 &=\;& -\frac{2}{3}\,{x^2\over b} - \frac{1}{3}\,{y^2\over c}\;.
\end{eqnarray}
To obtain the $K$-matrix unitarized amplitudes, the $\lambda_i$ are replaced by
\begin{equation}
\lambda_i\rightarrow\,\lambda_i^{\rm K} = {\lambda_i\over (1 -i\,\lambda_i)}\;,
\end{equation}
and the diagonal $t$-matrix is transformed back to the basis of physical states
with the aid of the matrix $U\,$.

\section{RESULTS}

Proceeding in the manner outlined in Sec.(\ref{sec:unit}), the unitarized
s-wave amplitudes for the production of $W_L$ and $Z_L^{\,0}$ pairs are
\begin{eqnarray}
a_0^{\rm K}(\gamma\,\gamma\rightarrow\,W_L^+\,W_L^-\,) &=\;& \frac{2}{3}\left(
{y\over 1 - i\,(c + \mathstrut\displaystyle {1\over 3}\,
\mathstrut\displaystyle{y^2\over c})} - {x\over 1 - i\,(b +
\mathstrut\displaystyle{2\over 3}\,\mathstrut\displaystyle{x^2\over b})}
\right)\;, \\
a_0^{\rm K}(\gamma\,\gamma\rightarrow\,Z_L^{\,0}\,Z_L^{\,0}\,) &=\;&
\frac{2}{3}
\left({y\over 1 - i\,(c + \mathstrut\displaystyle{1\over 3}\,
\mathstrut\displaystyle{y^2\over c})} + 2\,{x\over 1 - i\,(b +
\mathstrut\displaystyle{2\over 3}\,\mathstrut\displaystyle{x^2\over b})}
\right)\;,
\end{eqnarray}
where we have dropped terms of order $\alpha^3$, and removed the symmetry
factors associated with the identity of the $\gamma$'s and $Z_L^{\,0}$'s.
Taking
into account the definitions of $a\,$, $b\,$ and $d\,$, it can be seen that, at
the  Higgs pole, the unitarized amplitudes introduce a width which is equal to
the sum of $\Gamma(H\rightarrow\, W_L^+W_L^-)\,$, $\Gamma(H\rightarrow\,Z_L^{\,
0}Z_L^{\,0})\,$ and $\Gamma(H\rightarrow\,\gamma\,\gamma)$.

The cross sections obtained from these amplitudes are displayed in Figs.
(\ref{zz}-\ref{ww1}). For Higgs masses of .7 TeV and 1.0 TeV, the cross
sections
are in good agreement with those of Ref.\ \cite{boji}. This is consistent with
the situation found for the unitarized gauge boson scattering amplitudes
\cite{repsuch,dicrepkm}. When $m_H$ exceeds a few TeV, the unitarized
amplitudes
rise less rapidly than those presented in Ref.\ \cite{boji}, and the maximum
cross section for large $m_H$ is about a factor of three smaller. Fig.
(\ref{ww1}) shows the contribution of the one-loop correction to the cross
section for $\gamma\,\gamma\rightarrow\,W_L^+\,W_L^-\,$ scattering. A
comparison
of this figure with the result for the complete amplitude, Fig.(\ref{ww}),
shows
that the Born and one-loop amplitudes interfere substantially.

In conclusion, we find that unitarity corrections to the cross sections for
$\gamma\,\gamma\rightarrow\,Z_L^{\,0}\,Z_L^{\,0}\,$ and $\gamma\,\gamma
\rightarrow\,W_L^+W_L^-\,$ are large for $m_H\agt\,5\,$TeV. Nevertheless, in
the
large $m_H$ limit there is still a cross section on the order of several
femtobarns for either of these processes. This implies that searches for
ultraheavy fermions which make use of photon fusion \cite{chan} must take
into account the sources of gauge boson pairs discussed here.

\acknowledgments

One of us (WWR) wishes to thank the Center for Particle Physics at the
University of Texas for its hospitality during the initial stages of this work.
This research was supported in part by the National Science Foundation under
Grant 90--06117 and by the U. S. Department of Energy under Contract No.
DE--FG02--85ER40200.

\figure{The s-wave total cross section for $\gamma\,\gamma\rightarrow\,
Z_L^{\,0}\,Z_L^{\,0}\,$ is plotted for $m_H = .7,\,1.0\;{\rm and}\,10\,$TeV.
\label{zz}}
\figure{The s-wave total cross section for $\gamma\,\gamma\rightarrow\,
W_L^+\,W_L^-\,$ including the Born and one-loop contributions is plotted for
$m_H = .7,\,1.0\;{\rm and}\,10\,$TeV, with $\vert \cos\theta\vert\leq\,\cos(
\pi/6)$.\label{ww}}
\figure{Same as Fig.(\ref{ww}) except that only the one-loop contribution is
included.\label{ww1}}

\begin{references}
\bibitem{boji} E. E. Boss and G. V. Jikia, Phys.\ Lett.\ B {\bf 275}, 164
(1992).
\bibitem{hr-m} M. Herrero and E. Ruiz--Morales, Phys.\ Lett.\ B {\bf 296}, 397
(1992).
\bibitem{chan} M. S. Chanowitz, Phys.\ Rev.\ Lett. {\bf 69}, 2037, (1992).
\bibitem{repsuch} W. W. Repko and C. J. Suchyta III, Phys.\ Rev.\ Lett. {\bf
64}, 859 (1989).
\bibitem{dicrepkm} D. A. Dicus and W. W. Repko, Phys.\ Lett.\ B {\bf 228}, 503
(1989).
\bibitem{clt}J.\ M.\ Cornwall, D.\ N.\ Levin and G.\ Tiktopoulos, Phys.\ Rev.\
D
{\bf 10}, 1145 (1974).
\bibitem{lqt}B.\ W.\ Lee, C.\ Quigg, and H.\ Thacker, Phys.\ Rev.\ D {\bf 16},
1519 (1977).
\bibitem{cg}M.\ S.\ Chanowitz and M.\ K.\ Gaillard, Nucl.\ Phys.\ {\bf B261},
379 (1985).
\bibitem{f} We have checked that the inclusion of the $\,\varepsilon^2\,f\,$
term modifies only the eigenamplitude $\lambda_3\,$, and, therefore, does not
affect our results.
\end{references}
\end{document}